\documentclass{article}

\usepackage{microtype}
\usepackage{graphicx}
\usepackage{subcaption}
\usepackage{booktabs} % for professional tables

\usepackage{hyperref}

\newcommand{\ket}[1]{\left|#1\right\rangle}

\newcommand{\mat}[1]{\mathbf{#1}}

\usepackage{amsmath}
\usepackage{svg}
\svgsetup{inkscapelatex=false}
\DeclareMathOperator*{\argmin}{arg\,min}
\newcommand{\T}{\mathbf{\text{T}}}
\renewcommand{\H}{\mathbf{\text{H}}}
\newcommand{\I}{\mathbf{\text{I}}}
\renewcommand{\S}{\mathbf{\text{S}}}
\newcommand{\CX}{\mathbf{\text{CX}}}
\newcommand{\Tr}{\mathbf{\text{Tr}}}
\usepackage{booktabs}
\newcommand{\timeout}{\textsc{timeout}}
\newcommand{\na}{\textemdash}
\usepackage{tikz}
\usetikzlibrary{arrows.meta}
\usetikzlibrary{positioning}

\newcommand{\mlpnode}[2]{%
  \tikz[baseline=(base.base),x=1.1mm,y=1.0mm,line width=0.25pt,>=Latex]{
    \node[inner sep=0,outer sep=0] (base) at (0,0) {};
    \node[font=\scriptsize,anchor=east] at (-6,0) {#1};
    \node[font=\scriptsize,anchor=west] at (5,0) {#2};
    \draw[-Latex] (-6.2,0) -- (-4.2,0);
    \draw[-Latex] (4.2,0) -- (6.2,0);
    \foreach \y in {-1,1} {
      \filldraw[fill=blue!60,draw=black] (-2.7,\y) circle (0.5);
    }
    \foreach \y in {-2,0,2} {
      \filldraw[fill=orange!70,draw=black] (0,\y) circle (0.5);
    }
    \filldraw[fill=green!60,draw=black] (2.7,0) circle (0.5);
    \foreach \yin in {-1,1} {
      \foreach \yhid in {-2,0,2} {
        \draw (-2.7,\yin) -- (0,\yhid);
      }
    }
    \foreach \yhid in {-2,0,2} {
      \draw (0,\yhid) -- (2.7,0);
    }
  }%
}
\usepackage{stfloats}
\usepackage{siunitx}

\newcommand{\best}[1]{\bfseries #1}
\usepackage[preprint]{icml2026}

\usepackage{amsmath}
\usepackage{amssymb}
\usepackage{mathtools}
\usepackage{amsthm}

\usepackage[capitalize,noabbrev]{cleveref}

\crefname{equation}{Eq.}{Eqs.}
\crefname{figure}{Fig.}{Figs.}
\crefname{table}{Tab.}{Tabs.}
\crefname{section}{Sec.}{Secs.}
\crefname{algorithm}{Alg.}{Algs.}

\theoremstyle{plain}

\theoremstyle{definition}

\theoremstyle{remark}

\usepackage{enumitem}

\icmltitlerunning{Fast and Scalable Quantum Circuit Synthesis}

\begin{document}

\twocolumn[
  \icmltitle{Beyond Reinforcement Learning:\\ Fast and Scalable Quantum Circuit Synthesis}
  % Lightweight Zero-Shot Quantum Unitary Synthesis via Learning Minimum Description Length



  \icmlsetsymbol{equal}{*}

  \begin{icmlauthorlist}
    \icmlauthor{Lukas Thei{\ss}inger}{bonn}
    \icmlauthor{Thore Gerlach}{esa}
    \icmlauthor{David Berghaus}{iais}
    \icmlauthor{Christian Bauckhage}{bonn,iais}
  \end{icmlauthorlist}

  \icmlaffiliation{bonn}{University of Bonn}
  \icmlaffiliation{iais}{Fraunhofer IAIS}
  \icmlaffiliation{esa}{European
Space Agency, Advanced Concepts Team}

  \icmlcorrespondingauthor{Lukas Thei{\ss}inger}{lukas.theissinger@uni-bonn.de}

  \icmlkeywords{quantum computing, quantum circuit synthesis, minimum description length, unitary synthesis, scalability, stochastic beam search, zero-shot generalization, foundation model}

  \vskip 0.3in
]
\printAffiliationsAndNotice{}  % no special notice (required even if empty)

% Quantum unitary synthesis addresses the problem of translating abstract quantum algorithms into sequences of hardware-executable quantum gates. Solving this task exactly is infeasible in general due to the exponential growth of the underlying combinatorial search space. Existing approaches suffer from misaligned optimization objectives, substantial training costs and limited generalization across different qubit counts. We mitigate these limitations by using supervised learning to approximate the minimum description length of residual unitaries and combining this estimate with stochastic beam search to identify near optimal gate sequences. Our method relies on a lightweight model with zero-shot generalization, substantially reducing training overhead compared to prior baselines. Across multiple benchmarks, we achieve faster wall-clock synthesis times while exceeding state-of-the-art methods in terms of success rate for complex circuits.
\begin{abstract}
Quantum unitary synthesis addresses the problem of translating abstract quantum algorithms into sequences of hardware-executable quantum gates.
Solving this task exactly is infeasible in general due to the exponential growth of the underlying combinatorial search space.
Existing approaches suffer from misaligned optimization objectives, substantial training costs and limited generalization across different qubit counts.
We mitigate these limitations by using supervised learning to approximate the minimum description length of residual unitaries and combining this estimate with stochastic beam search to identify near optimal gate sequences.
Our method relies on a lightweight model with zero-shot generalization, substantially reducing training overhead compared to prior baselines.
Across multiple benchmarks, we achieve faster wall-clock synthesis times while exceeding state-of-the-art methods in terms of success rate for complex circuits.
\end{abstract}

\section{Introduction}

Quantum computing is compelling in principle because quantum algorithms exploit superposition and interference to implement transformations that appear qualitatively inaccessible to classical computation~\cite{nielsen2010qcqi,montanaro2016quantum,shor1994algorithms,grover1996search}.
In practice, those algorithmic advantages are realized through unitary operations: a quantum algorithm is, at its core, a sequence of unitary matrices. This makes a simple but often under-emphasized point unavoidable: progress in quantum algorithms is bottlenecked not only by hardware, but by our ability to construct circuits that implement the right unitaries with realistic resource costs~\cite{nielsen2010qcqi}.
The task of finding a sequence of gates to build a quantum circuit which realizes a given unitary is coined Quantum Unitary Synthesis~(QUS)~\cite{shende2005synthesis}.

QUS closely parallels symbolic regression~\cite{augusto2000symbolic}. In both cases, the target is observed only through numerical representations (i.e., matrix elements of a unitary or floating-point evaluations of a function) while the objective is to recover a compact description composed from a finite set of primitives. This task goes beyond numerical approximation: distinct symbolic structures may induce similar numerical behavior, whereas small symbolic errors can lead to large numerical discrepancies.
This suggests that QUS, like symbolic regression, is fundamentally a search over discrete symbolic spaces rather than continuous optimization in isolation.

Classical approaches to QUS rely on heuristic search~\cite{paradis2024synthetiq}, exact optimization~\cite{nagarajan2025provably} or hand-crafted algebraic rules~\cite{gheorghiu2022t} and have proven effective only in limited regimes. As system size grows, the synthesis problem becomes dominated by a combinatorial explosion of possible gate sequences, while commonly used numerical objectives provide weak guidance: numerical proximity between unitaries need not reflect symbolic similarity and locally optimal choices obstruct globally correct constructions.
State-of-the-art supervised learning~(SL) methods~\cite{furrutter2024quantum,barta2025leveraging,chen2025uditqc} face similar issues.

\begin{figure*}[htbp]
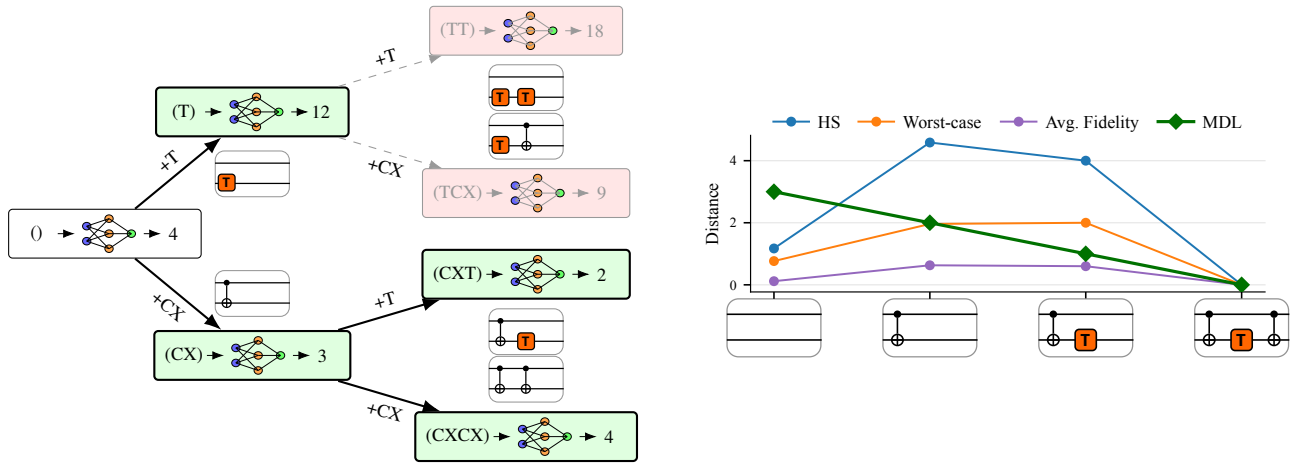

  \centering
  \begin{minipage}[c]{0.48\textwidth}
  \label{fig:Method}
    \centering
       \begin{tikzpicture}[
  x=2.8cm,
  y=1.8cm,
  >=Latex,
  node/.style={rectangle,rounded corners=1.5pt,draw,minimum width=6mm,minimum height=5mm,inner sep=2pt,font=\small},
  best/.style={node,fill=green!12,thick},
  pruned/.style={node,draw=black!40,text=black!40,fill=red!10},
  edge/.style={-Latex,thick},
  pruned edge/.style={-Latex,draw=black!40,dashed},
  label/.style={font=\footnotesize},
  edge label/.style={font=\scriptsize,fill=white,inner sep=1pt,sloped}
]
  \node[node] (s) at (0,0) {\mlpnode{()}{4}};

  \node[best] (a) at (0.7,0.9) {\mlpnode{(T)}{12}};
  \node (fig) [below=0.5mm of a] {
  \includesvg[width=1cm]{figures/t}};
  \node[best] (b) at (0.7,-0.9) {\mlpnode{(CX)}{3}};
  \node (fig) [above=0.5mm of b] {
  \includesvg[width=1cm]{figures/cx}};
  
  \node[pruned] (d) at (2,1.5) {\mlpnode{(TT)}{18}};
  \node (fig) [below=-0.1mm of d] {
  \includesvg[width=1cm]{figures/t_t}};
  \node[pruned] (e) at (2,0.3) {\mlpnode{(TCX)}{9}};
  \node (fig) [above=-0.1mm of e] {
  \includesvg[width=1cm]{figures/t_cx}};
  \node[best] (f) at (2,-0.3) {\mlpnode{(CXT)}{2}};
  \node[best] (g) at (2,-1.5) {\mlpnode{(CXCX)}{4}};
  \node (fig) [below=-0.1mm of f] {
  \includesvg[width=1cm]{figures/cx_t}};
  \node (fig) [above=-0.1mm of g] {
  \includesvg[width=1cm]{figures/cx_cx}};
  \draw[edge] (s) -- node[edge label,midway,above=2pt] {+T} (a);
  \draw[edge] (s) -- node[edge label,midway,below=2pt] {+CX} (b);

  \draw[pruned edge] (a) -- node[edge label,midway,above=2pt] {+T} (d);
  \draw[pruned edge] (a) -- node[edge label,midway,below=2pt] {+CX} (e);
  \draw[edge] (b) -- node[edge label,midway,above=2pt] {+T} (f);
  \draw[edge] (b) -- node[edge label,midway,below=2pt] {+CX} (g);

\end{tikzpicture}
  \end{minipage}\hfill
  \begin{minipage}[c]{0.47\textwidth}
  \label{fig:teaser}
    \centering
    \includesvg[width=\linewidth]{figures/distances}
  \end{minipage}

  \caption{\textbf{Fast and Scalable Quantum Circuit Synthesis.}
  \textbf{Left:} Illustration of our synthesis search. We use an MDL predictor with beam search at inference time: candidate circuits are expanded by appending gates (+$\T$, +$\CX$). Each node denotes a partial gate sequence ($\T$, $\CX$, $\T\CX$, $\CX\T$, $\T\T$, $\CX\CX$) and the numbers show the predicted remaining MDL. Green nodes are kept, while red nodes are pruned from the search.
  \textbf{Right:} Comparison of several distance measures: Hilbert-Schmidt (HS), worst-case, average fidelity and MDL evaluated on representative two-qubit circuits, illustrating how the choice of metric changes the landscape as circuits approach the target.}
  \label{fig:overview}
\end{figure*}

To address this mismatch,~\cite{yu2025mdlformer} propose using the Minimum Description Length~(MDL)~\cite{kolmogorov1963tables} as a structurally meaningful objective.
It is defined as the smallest number of symbols required to represent an expression.
The MDL decreases monotonically along correct symbolic sequences and thus provides guidance for exploration in discrete symbolic spaces, in opposition to conventional norm-based distance metrics (see~\cref{fig:overview}).
Recent methods based on reinforcement learning~(RL) attempt to utilize this by designing reward functions to reflect symbolic similarity~\cite{rietsch2024unitary,kremer2025optimizing}.
However, these models require long training times and exhibit limited generalization across different qubit counts.

% Existing MDL-based approaches estimate this quantity via supervised learning on synthetic data, where ground-truth expressions---and hence their MDL---are known by construction, avoiding reinforcement learning but relying on generalization from synthetic symbolic distributions to practical synthesis settings.

% We build upon recent advances in the symbolic regression community, namely \mdlf \cite{yu2025mdlformer} which achieved a significant improvement in the recovery rate of symbolic expressions. While previous works in symbolic regression typically tried to minimize some vector-norm based loss between the observations and the reconstruction, \cite{yu2025mdlformer} argued that this is inherently flawed because two expressions might be very close in \emph{symbolic space} (for example just missing one exponentiation) but can have very different plots and hence high reconstruction losses (we observed analogous behavior for unitary-synthesis, see~\cref{fig:teaser}). Consequently, even a model predicting the correct sequence of symbols in an autoregressive fashion might receive bad feedback for some intermediate decisions, because those intermediate decisions may temporarily increase the reconstruction loss. Moreover, two expressions might have similar reconstructions but are inherently different. \cite{yu2025mdlformer} therefore propose to work with the Minimum Description Length (MDL) which denotes the smallest number of symbols that is needed to express an expression. 

In this work, we propose an RL-free approach to QUS that leverages the efficiency of SL while adapting the MDL framework to overcome the mismatch between symbolic structure and reconstruction error.
We generate synthetic training data by sampling random quantum circuits and their corresponding unitaries, enabling supervised training to predict the MDL of candidate partial circuits.
The trained model is then used as a value function within a beam-search procedure to efficiently explore the combinatorial space of circuit decompositions.
An overview of our method is found in~\cref{fig:overview} and
our contributions are summarized as follows:
\begin{itemize}[leftmargin=2em]
    \item \textbf{Synthesis via MDL}: We formulate quantum circuit synthesis as the problem of estimating the remaining optimal gate cost of a residual target unitary using the MDL, yielding a structurally meaningful value function for guiding symbolic search.
    \item \textbf{Lightweight Model}: We find that a lightweight multi-layer perceptron achieves accuracy better than a transformer architecture \cite{vaswani2017attention} which one might expect to be more expressive (\cref{sec:transformer}), resulting in fast inference. Training time is largely reduced compared to state-of-the-art RL approaches~\cite{rietsch2024unitary,kremer2025optimizing}.
    \item \textbf{Zero-shot Capabilities}: We train a single model on a broad synthetic distribution and deploy it \emph{zero-shot}, without task-specific retraining, across all evaluation settings. Consistent with prior findings~\cite{fim_sde,fim_pp}, this enables effective generalization-including to circuits with varying numbers of qubits-while avoiding the costly per-qubit training required by previous approaches.
    \item \textbf{State-of-the-Art Performance}: We show that our method is state-of-the-art in both time and recovery rate by comparing it against previous classical and RL-based baselines on a variety of established benchmarks~\cite{lu2023qasbench}.
\end{itemize}

\section{Related Work}\label{sec:related}
A large body of work focuses on classical optimization for quantum circuit synthesis with arbitrary gate sets, exploiting the maturity of general-purpose solvers.
\texttt{Synthetiq}~\cite{paradis2024synthetiq} utilizes simulated annealing~\cite{kirkpatrick1983optimization} to iteratively refine a randomly built circuit until a certain criterion is met.
\texttt{QuantumCircuitOpt}~\cite{nagarajan2021quantumcircuitopt,nagarajan2025provably} converts the exact circuit problem into a mixed-integer linear program, enabling the use of sophisticated branch-and-bound solvers such as \texttt{Gurobi}~\cite{gurobi}.
Automated reasoning is used in~\cite{zak2025reducing} by casting unitary synthesis as a model counting problem.
While these approaches provide strong optimality guarantees and broad expressivity, their reliance on general-purpose solvers often limits scalability as circuit size and search space grow.

For Clifford+T synthesis, strong guarantees are provided by exploiting the algebraic structure of the gate set.
\cite{kliuchnikov2013fast} showed that single-qubit unitaries admit exact, ancilla-free decompositions and described an efficient construction strategy for recovering the corresponding circuit.
This result was extended to the multi-qubit regime, presenting constructive algorithms that match theoretical guarantees~\cite{giles2013exact,gheorghiu2022t}.
A more efficient heuristic for synthesizing a T-count optimal unitary was developed in~\cite{mosca2021polynomial}.
However, these algorithms exhibit exponential runtime in the worst case, restricting their practical use to small instances.
The underlying number-theoretic structure nevertheless motivates our focus on gate count-aware search over the same gate family.

Recent progress in unitary synthesis complements exact methods with data-driven heuristics, reflecting a broader shift away from purely symbolic and heuristic techniques~\cite{wang2022quantumnas}.
The most prominent supervised methods rely on generative diffusion models~\cite{furrutter2024quantum,chen2025uditqc,barta2025leveraging}.
% Recent progress on quantum architecture search complements exact methods with data-driven heuristics.
% QuantumNAS and related approaches explore noise-aware design spaces by combining neural surrogates with guided search, demonstrating that lightweight models can steer exploration under tight hardware constraints~\cite{wang2022quantumnas}.

QAS-Bench further systematizes this perspective by providing a reusable benchmark suite that exposes dozens of parameterized search tasks, enabling controlled comparisons between competing heuristics~\cite{lu2023qasbench}.
Our beam-search blueprint borrows the benchmarking philosophy (fixed search budgets, standardized action sets) and adapts it to the Clifford+T domain.

Owing to the vast design space of quantum circuits, several recent works have investigated the automation of unitary synthesis using reinforcement learning (RL)~\cite{kundu2024kanqas,kremer2024practical,zen2025quantum,kundu2025tensorrl, rietsch2024unitary}.
Building on the tree-search–based AlphaZero algorithm~\cite{silver2018general}, these approaches learn value functions that guide the search toward promising unitary synthesis strategies~\cite{rietsch2024unitary}.
These methods have been extended to dynamic circuits~\cite{valcarce2025unitary} and the Pauli channel representation~\cite{kremer2025optimizing}.
While these methods demonstrate substantial improvements over unguided search, they remain restricted to a fixed number of qubits.
We pursue a complementary direction in which a compact neural network produces gate count estimates that bias stochastic beam expansions, yielding a simple yet effective heuristic.

\section{Background}\label{sec:background}

We begin with a brief review of circuit-based quantum computing and \emph{quantum unitary synthesis}~(QUS), which provides the background needed for the remainder of the paper. We refer the interested reader to~\cite{nielsen2010qcqi} for a detailed introduction. 
% Further, we introduce the connection of quantum algorithms to the \emph{minimum description length} (MDL) view that underlies our proposed method.
Further, we connect the task of QUS with \emph{Markov decision processes}~(MDP).

\textbf{Quantum Computing}
Quantum computing relies on \emph{qubits} for computation. They are physical objects and therefore behave according to the laws of quantum mechanics.
An $n$-qubit quantum state $\ket{\psi}$ is a unit-norm vector in a $2^n$-dimensional complex Hilbert space; this encodes an exponentially large state space and computation is performed by applying unitary transformations (matrices) called \emph{quantum gates}.
A quantum algorithm consists of a sequence of such gates followed by a measurement, which probabilistically collapses the quantum state to a classical outcome.
While the full quantum state is not directly observable, measurements provide limited information about it. By choosing the gate sequence appropriately, a quantum algorithm can amplify the probability of desired outcomes through constructive interference.

% An $n$-qubit pure state is a unit vector $\ket{\psi}\in\mathbb{C}^{2^n}$.
% Quantum circuits implement unitary transformations $\mat U\in \mathrm{U}(2^n)$.
% Because global phase is physically unobservable, we consider $\mat U$ and $e^{i\phi}\mat U$ equivalent and define exact synthesis up to global phase.

\textbf{Quantum Unitary Synthesis}
In the circuit-model, quantum algorithms are represented by quantum circuits. A quantum circuit is a sequence $C=(\mat G_1,\dots,\mat G_m)$ with $\mat G_i\in\mathbb C^{2^n\times 2^n}$ and its implemented unitary is $\mat U(C) \;=\; \mat G_m \cdots \mat G_1$ (notice that the order gets reversed).
QUS concerns the construction of a circuit $C$ that implements a target unitary operation $\mat U^\star$ under architectural and resource constraints. Its objective is given by:
\begin{equation*}
    \argmin_{C=(\mat G_1,\dots,\mat G_m)} d\left(\mat U(C),\mat U^\star\right),\quad \mat G_i\in\mathcal G,
\end{equation*}
where $d$ is a suitable objective function on unitary operators and $\mathcal G$ is a specific quantum gate set.

%\todo[inline]{Separate equations take a lot of space, why mention all of them?}
Common choices for $d$ are distances between unitary matrices. Following \cite{nielsen2010qcqi}, we quantify the (worst-case) unitary implementation error by
\[
d_{worst}(U,V)\;=\;\max_{\ket{\psi}:\,\|\psi\|_2=1}\|(U-V)\ket{\psi}\|_{2}
\]
and the Hilbert-Schmidt distance
\[
d_{\mathrm{HS}}(U,V)\;=\;\|U-V\|_{F},
\]
which is conveniently computable \cite{gilchrist_distance_2005,watrous_tqi_2018}.
A widely used alternative is to optimize a fidelity metric
\begin{equation} \label{eq:AvgFidelity}
F_{\mathrm{avg}}(U,V)\;=\;\frac{|\Tr(U^\dagger V)|^2 + D}{D(D+1)},\qquad D=2^n,    
\end{equation}
which depends on the trace overlap and has a direct operational interpretation \cite{nielsen_simple_2002,gilchrist_distance_2005}: it equals the expected state overlap between $\mathbf{U}\ket{\psi}$ and $\mathbf{V}\ket{\psi}$ when $\ket{\psi}$ is drawn uniformly at random from pure states~\cite{nielsen_simple_2002,nielsen2010qcqi}. A comparison of these distance functions on a sample synthesis path can be seen in~\cref{fig:overview}.

\textbf{Considered Gate Set}
We focus on the Clifford+T family because it is universal, heavily used in fault-tolerant implementations and induces a discrete search space amenable to combinatorial optimization \cite{bravyi2005universal,kliuchnikov2013fast,amy2013mitm}.
% A common generating set is $\{H,S,T,\mathrm{CNOT}\}$, where the Clifford subgroup is generated by $\{H,S,\mathrm{CNOT}\}$ and $\T$ is a non-Clifford gate.
Clifford gates enable efficient classical simulation, while adding $\T$ gates yields universality of quantum computations.
Although fault-tolerant cost is often dominated by $\T$ gates, our method only assumes a discrete library $\mathcal{G}$ and a chosen description length; here we instantiate it with gate count.

%QUS is naturally expressed in terms of an MDP.
%Given a partial circuit prefix $C_{1:t}$ with unitary $\mat U_{1:t}=\mat U(C_{1:t})$, define the residual unitary as 
%\begin{equation} \label{eq:residual}
%\mat R_t \;=\; \mat U_{1:t}^\dagger \mat U^\star,
%\end{equation}
%which corresponds to the state at time $t$.
%Taking an action (applying a gate $\mat G\in\mathcal G$) leads to a new state $\mat R_{t+1}$, which incurs a reward $r(\mat R_t,\mat G)$.
%The goal is to maximize the cumulative reward
%\todo[inline]{There is a slight mistake here:\sum_{\mat G\in C}r(\mat R_t,\mat G) should be \sum_{t = 0}^{|C|-1} r(\mat R_t,\mat G_t), because otherwise the state would be stationary, which it obviously is not}
%\begin{align}
%    \max_{C=(\mat G_1,\dots,\mat G_T)}V^C(\mat R_1),\ V^C(\mat R_t)=\sum_{\mat G\in C}r(\mat R_t,\mat G), \label{eq:max_cum_reward}
%\end{align}
%with the termination criterion $T=\min\{t: d\left(\mat R_t,\mat I\right)\le \epsilon\}$.
%In~\cite{rietsch2024unitary,valcarce2025unitary,kremer2025optimizing}, the reward function is chosen as $-1$ in every step until termination upon exact synthesis. 
\textbf{Decision Process Formulation}
QUS is naturally expressed in terms of an MDP.
Given a partial circuit prefix $C_{1:t}$ with unitary $\mat U_{1:t}=\mat U(C_{1:t})$, define the residual unitary as
\begin{equation} \label{eq:residual}
\mat R_t \;=\; \mat U_{1:t}^\dagger \mat U^\star,
\end{equation}
which corresponds to the state at time $t$.
Taking an action (applying a gate $\mat G\in\mathcal G$) leads to a new state
$\mat R_{t+1} \;=\; \mat G^\dagger \mat R_t$, and incurs a reward
\begin{equation*}
r(\mat R_t,\mat G)=
\begin{cases}
0, & d(\mat R_t,\mat I)\le \epsilon,\\
-1, & \text{otherwise}.
\end{cases}
\end{equation*}
For a fixed circuit $C=(\mat G_1,\dots,\mat G_T)$, the cumulative return is
\begin{equation} \label{eq:max_cum_reward}
V^C(\mat R_0)=\sum_{t=0}^{T-1} r(\mat R_t,\mat G_{t+1}),
\end{equation}
with termination at $T=\min\{t: d(\mat R_t,\mat I)\le \epsilon\}$.
% The corresponding Bellman optimality equation is
% \begin{equation}
% V^*(\mat R)=\max_{\mat G\in\mathcal G}\Big[r(\mat R,\mat G)+V^*(\mat G^\dagger \mat R)\Big],
% \end{equation}
% with $V^*(\mat R)=0$ when $d(\mat R,\mat I)\le \epsilon$.

Under this reward, the optimal return equals the negative minimum remaining gate count, so
\begin{equation*}
-\,V^*(\mat R)=\mathrm{MDL}_{\mathcal G}(\mat R).
\end{equation*}

Obtaining an optimal solution (maximum cumulative reward in~\cref{eq:max_cum_reward}) boils down to finding the best path in the graph created by all possible gate applications.
This graph is exponentially large, since every $k$-qubit gate has $\binom{n}{k}$ possibilities to be applied at every step.
Therefore, direct materialization is infeasible and practical methods resort to guided exploration mechanisms, notably RL-based frameworks.
% Completing the circuit is equivalent to synthesizing $\mat R_i$ (up to global phase) and the remaining optimal work is
% \[
% \mathrm{MDL}_{\mathcal{G}}(\mat R_i),
% \]
% i.e., the minimum number of additional gates required to finish the synthesis.

Instead of relying on the recursive Bellman equation used in RL, our approach is RL-free and learns a predictor of the residual unitary's MDL for approximating the value function and uses it to prioritize expansions in beam search.

\section{Methodology}\label{sec:methodology}

% \paragraph{Minimum Description Length}

% We can view a circuit as a discrete \emph{description} of the unitary under the language induced by $\mathcal{G}$ and define the MDL of a unitary as the length of its shortest exact description:
% \[
% \mathrm{MDL}_{\mathcal{G}}(\mat U) \;\triangleq\; \min_{C:\; U(C)=\mat U} |C|.
% \]
% Thus, in our setting, \textbf{MDL is exactly the minimum gate count} required to represent $\mat U$ over $\mathcal{G}$ (up to global phase).
% This framing is useful because $\mathrm{MDL}_{\mathcal{G}}(\mat U)$ is a natural \emph{cost-to-go} quantity for search, even though computing it exactly is itself hard.

% We start by reformulating the unitary synthesis problem in terms of MDL.
% We fix a discrete gate library $\mathcal{G}$ (Clifford+$\T$) and represent a quantum circuit as a finite sequence of gates $C=(g_1,\dots,g_L)$ with $g_\ell\in\mathcal{G}$ as seen before.
To connect synthesis to MDL, we view a circuit as a discrete description string over the alphabet $\mathcal{G}$.
Let $\ell(\mat G)\ge 0$ denote the (fixed) expression-length assigned to gate $\mat G$ under a chosen encoding and define the description length of a circuit as
\[
\ell(C) \;\triangleq\; \sum_{i\le m} \ell(\mat G_i).
\]
The MDL of a unitary under $\mathcal{G}$ is the length of its shortest description
\[
\mathrm{MDL}_{\mathcal{G}}(\mat U) \;\triangleq\; \min_{C:\;d\left(\mat U(C),\mat U\right)\le \epsilon}\; \ell(C).
\]
We choose the simple instantiation $\ell(\mat G)\equiv 1$, from which follows $\ell(C)=|C|$ and $\mathrm{MDL}_{\mathcal{G}}(\mat U)$ is exactly the minimum gate count needed to represent $ \mat U$ over $\mathcal{G}$.

A convenient state representation for guided search is the residual unitary induced by a partial circuit as defined by (\cref{eq:residual}).
Completing the synthesis from the current prefix is equivalent to exactly synthesizing $\mat R_t$ and the remaining optimal future cumulative reward is precisely the MDL.
% \[
% \mathrm{MDL}_{\mathcal{G}}(\mat R_\ell).
% \]
In other words, $\mathrm{MDL}_{\mathcal{G}}(\mat R_t)$ is an ideal cost-to-go: a perfect heuristic would predict how many additional gates are required from the current search state. Our method operationalizes this MDL view by learning to predict $\mathrm{MDL}_{\mathcal{G}}(\mat R_t)$ and using that prediction to prioritize expansions in the search.

\subsection{Predicting the Minimum Description Length}
We train the MDL predictor in a supervised fashion, which requires pairs of residual unitaries and their remaining MDL. Since computing the exact
$\mathrm{MDL}_{\mathcal{G}}(\cdot)$ is computationally hard, we generate \emph{consistent} labels from heuristically optimized circuits and treat the resulting targets as an approximation to MDL that is sufficient for guiding the search.

\textbf{Data Generation} \label{p:data}
We construct training circuits via rejection sampling to control circuit difficulty and keep the marginal distribution over target $\T$-count approximately uniform. Concretely, we draw a target $\T$-count $k\in[0,20]$ uniformly at random and propose a random Clifford+$\T$ circuit by (i) sampling a random Clifford circuit, choosing Clifford gates and their acted-on qubits uniformly at random, (ii) inserting exactly $k$ $\T$-gates at uniformly sampled positions and (iii) randomly shuffling the resulting gate sequence. We then apply a lightweight peephole optimizer (a fixed set of local rewrite rules) to remove obvious cancellations and normalize the circuit; if the optimizer changes the $\T$-count, we reject the proposal and resample. 

We use a lightweight curriculum to better train on high $\T$-count circuits. Let $C$ be an accepted optimized circuit with unitary $\mat U^\star=\mat U(C)$ and gate count $|C|$. We form multiple supervised examples by choosing cut positions $t\in\{0,\dots,|C|\}$ and taking the prefix $C_{1:t}$ with unitary $\mat U_{1:t}=\mat U(C_{1:t})$. The search state is the residual unitary
\[
\mat R_t \;=\; \mat U_{1:t}^\dagger \mat U^\star,
\]
i.e., the unitary remaining after committing to the prefix. The ideal target is then $\mathrm{MDL}_{\mathcal{G}}(\mat R_t)$. Since $C_{t+1:|C|}$ implements $\mat R_t$, we optionally re-run the peephole optimizer on this suffix to obtain a shorter exact description $\widetilde{C}_t$ and set the label $y_t = |\widetilde{C}_t|$. Under $\ell(\mat G)\equiv 1$, $y_t$ proxies the remaining minimum gate count.

To emphasize hard (long-residual) states, we do not sample $t$ uniformly. Instead, we align cuts with non-Clifford structure: letting $k'$ be the number of $T$ gates in $C$, if $k'\ge 5$ we include a cut after the $\lfloor k'/2\rfloor$-th $T$ gate, and if $k'\ge 10$ we additionally cut after the $\lfloor 3k'/4\rfloor$-th $T$ gate (in circuit order). This broadens the label distribution and upweights high-$T$ circuits.

\textbf{Model Input}
Each residual unitary $\mat R_t$ is converted into the network input by removing an arbitrary global phase and stacking real and imaginary parts into a real-valued tensor (we discuss alternative tokenization methods in appendix \ref{sec:transformer}).

We remove the arbitrary global phase of each residual unitary in the following way:
Given a residual $\mat R$, we scan its entries in a fixed order (row-major) and select the first index $(i^\star,j^\star)$ such that $|R_{i^\star j^\star}|>\varepsilon$ for a small threshold $\varepsilon$ close to $0$.
Let $z = R_{i^\star j^\star} = r \exp{\theta i}$, where $r$, $\theta$ is the polar form of the complex number $z$.
We then phase-normalize by multiplying the entire matrix by $e^{-i\theta}$, resulting in $\widehat{ \mat R} \;=\; e^{-i\theta} \mat R$.
By construction, the reference entry $\widehat{R}_{i^\star j^\star}$ is real and nonnegative and the procedure is invariant to global phase:
$\widehat{ \mat R}$ is unchanged if $\mat R$ is replaced by $e^{i\phi} \mat R$.
This normalization is necessary because, in physics, a global phase $e^{i\phi}$ is unobservable, so there is no way to distinguish $\mat R$ from $e^{i\phi}\mat R$.

\textbf{Loss Function}
Given a phase-normalized residual $\widehat{\mat R}_t$, the network $f_\theta$ produces a single scalar prediction
$\hat y_t = f_\theta(\widehat{\mat R}_t)$, intended to estimate the remaining description length $y_t = |\widetilde C_t|$. We train $f_\theta$ by standard squared-error regression:
\[
\mathcal{L}(\theta)
\;=\;
\mathbb{E}_{(\widehat{\mat R}_t,\, y_t)\sim \mathcal{D}}
\bigl[\bigl(f_\theta(\widehat{\mat R}_t)-y_t\bigr)^2\bigr],
\]
implemented as the mean squared error over minibatches. At test time, $-\hat y_t$ is used as an estimate
of the value function.

%\subsection{Noisy Data}
%The optimizer does not give perfect target labels, as generating true circuit depth labels is as hard as solving our actual problem, i.e. NP-hard \cite{vanDeWetering2023hardness}.
%Therefore, our data is considered noisy. The data is noisy, because the greedy peephole optimizer does not guarantee to find the optimal solution. However, we do not need the optimal depth, we need to decide which gate to add next. This has the following consequence:
%\begin{enumerate}
%    \item An overall bias in the data is not a problem, because we only care about relative differences.
%    \item As long as our optimization routine ranks short circuits better than long circuits, we are fine.
%    \item Near-ties can be mis-ranked.
%\end{enumerate}
%We figured that out by trying to use RL fine tuning to boost the models performance, but we found that it actually decreases performance. Hinting that the new samples had a different bias to them, which did not match the generated training data.

\subsection{Inference with Stochastic Beam Search}\label{sec:inference}

At inference time we are given a target unitary $\mat U^\star$ and a trained predictor $f_\theta(\cdot)$ that estimates the MDL, i.e., the remaining minimum gate count.
If $f_\theta$ were perfect, repeatedly choosing the next gate that minimizes the predicted remaining MDL would recover an optimal circuit.
In practice the predictor is imperfect (both due to approximation error and because our labels are proxy MDL values), so purely greedy decoding is brittle.
We therefore use a width-bounded search that mixes exploitation with controlled exploration while retaining high throughput.

\begin{algorithm}[htbp]
\small
\caption{Inference}\label{alg:inference}
\begin{algorithmic}[1]
\REQUIRE Target unitary $
\mat U^\star$, gate set $\mathcal{G}$, MDL predictor $f_\theta$, beam width $B$, max steps $T$, temperature $\tau$, goal tolerance $\varepsilon$
\ENSURE A circuit $C$
\STATE $\mat R_0 \gets \mat U^\star$;\;\; $\text{beam} \gets \{(\mat R_0, \langle\rangle)\}$;\;\; $\text{sol} \gets \emptyset$
\FOR{$t = 0$ to $T-1$}
    \STATE $\text{cand} \gets \emptyset$
    \FORALL{$(\mat R_t,C_t)$ in beam}
        \IF{$d(\mat R_t, \mat I) \leq \varepsilon$}
            \STATE $\text{sol} \gets \text{sol} \cup \{(C_t,t)\}$
            \STATE \textbf{continue}
        \ENDIF
        \FORALL{$\mat G \in \mathcal{G}$}
            \STATE $\mat R_{t+1} \gets \mat G^\dagger \mat R_t$
            \STATE $C_{t+1} \gets C_t \mathbin{\Vert} \mat G$
            \STATE $\text{mdl} \gets - f_\theta(\mat R_{t+1})$
            \STATE $\text{cand} \gets \text{cand} \cup \{(\mat R_{t+1},C_{t+1},\text{mdl})\}$
        \ENDFOR
    \ENDFOR
    \IF{$\text{sol}\neq \emptyset$}
        \STATE \textbf{return} best $C$ in $\text{sol}$
    \ENDIF

    \IF{$\text{cand}=\emptyset$} \STATE \textbf{break} \ENDIF
    
    \STATE $\,\text{beam} \gets \text{Top-}B\big(\text{cand},\; \text{mdl}/\tau + \mathrm{Gumbel}(0,1)\big)$
\ENDFOR
\end{algorithmic}
\end{algorithm}

% We formulate synthesis as searching over residual unitaries.
% Starting from $\mat R_0 = \mat U^\star$, each action applies a gate $g\in\mathcal{G}$ to reduce the residual, e.g.,
% \[
% \mat R_{\ell + 1} \;=\; g^\dagger \mat R_\ell,
% \]
% so that reaching the identity (up to global phase) corresponds to completing the synthesis.
% A partial trajectory induces a gate sequence (constructed in reverse order under the above update) and a residual $\mat R_\ell$; the remaining optimal work from this state is $\mathrm{MDL}_\mathcal{G}(\mat R_\ell)$, which we approximate with $f_\theta(\mat R_\ell)$.

\textbf{Approximate Optimal Value Function}
To rank candidates during beam search, we approximate the optimal value function by the negative predicted MDL
\[
V^*(\mat R_t)\approx- f_\theta(\mat R_t).
\]
Beam search maintains the best $B$ residuals under this score, expands each residual by all allowed gate applications, evaluates $f_\theta$ on the resulting residuals in a single batch and then selects the next beam.

\textbf{Stochastic Selection}
To avoid over-committing to the model early, we use a stochastic selection rule based on Gumbel-top-$B$ sampling~\cite{danihelka2022policy}: we add i.i.d.\ Gumbel noise to the negative scores and take the top-$B$.
This is equivalent to sampling without replacement from a softmax distribution over $- f_\theta$ with a softness-controlling temperature $\tau$:
\begin{align*}
    \mathrm{arg\,top}_{\mat G}\left(-f_\theta(\mat G^\dagger \mat R)/\tau+g, B\right),\quad g\sim \mathrm{Gumbel}(0,1).
\end{align*}
This yields a simple exploration-exploitation tradeoff with minimal engineering overhead and preserves the key advantage of beam search in our setting: the entire expansion-and-scoring step is completely parallel across candidates.

\textbf{Termination Criterion}
We declare convergence once the iterate is $\varepsilon$-close to the identity under a
phase-invariant notion of discrepancy, i.e., $d(\mathbf{R}_t,\mathbf{I}) \le \varepsilon$.
We measure this closeness using the average gate fidelity~\cite{nielsen_simple_2002}, defined in \cref{eq:AvgFidelity}.

Finally, we run multiple independent trials and keep the best solution.
Each trial corresponds to an independent stochastic run (different Gumbel noise) and can also apply symmetry transformations that preserve synthesis difficulty but change the search landscape.
In particular, permuting qubits via a permutation matrix $\mat P$ yields the equivalent target $\mat U^\star_\pi = \mat P \mat U^\star \mat P^\dagger$; a circuit for $\mat U^\star_\pi$ is converted back by relabeling the corresponding qubit indices.
When the action set is closed under adjoints, we additionally allow inverse trials that synthesize ${\mat U_\pi^\star}^\dagger$ and converts the result to a circuit for $\mat U^\star$ by reversing the gate sequence and taking adjoints.
In all cases, trials are independent and parallelizable. A pseudo code of our inference method can be found in \cref{alg:inference}.

\section{Experiments}\label{sec:experiments}

\begin{figure*}[ht]
\centering
    \begin{subfigure}[t]{0.48\textwidth}
    \centering
    \includesvg[width=\linewidth]{gfx/histogram_comparison_4}
    \caption{4 qubit comparison}
    \end{subfigure}
    \begin{subfigure}[t]{0.48\textwidth}
    \centering
    \includesvg[width=\linewidth]{gfx/histogram_comparison_5}
    \caption{5 qubit comparison}
    \end{subfigure}
    \caption[Comparison to RL baseline]{Success counts (out of 100 targets per $\T$-count, higher is better) for MDL-guided beam search versus the RL baseline of~\cite{rietsch2024unitary} and annealing algorithm of~\cite{paradis2024synthetiq} on 4 and 5-qubit instances. Our method uses beam width $B{=}10$ and $8000$ trials per instance (avg.\ $\sim$22s runtime) and declares success when $F_{\mathrm{avg}}(\mat U(C),\mat U^\star)\ge 0.9$ (\cref{eq:AvgFidelity}).
    Results for RL at high T-counts are unavailable because they are not reported in~\cite{rietsch2024unitary}, likely due to the associated computational cost.
    }
    \label{fig:Comparison}
\end{figure*}

%\todo[inline]{L: Still not exactly 8 pages}
We trained one MDL predictor on $n=5$ qubits once for all further experiments.
The MDL-predictor is a feed-forward multilayer perceptron with hidden dimensions $(1024,512,128)$ and a softplus output to ensure non-negative predictions.  Training the MLP takes approximately $6$ hours in our implementation and is dominated by data generation and unitary construction; we use 30 CPU cores and a 4 \textsc{GB} GPU. This is substantially cheaper than RL training regime reported in~\cite{rietsch2024unitary} (7 days of training). We also trained a transformer model with different tokenization strategies, yielding in no improvement over the MLP (see~\cref{sec:transformer}).

To apply the MDL predictor to a different number $m<5$ of qubits, we embed an $m$-qubit target unitary $\mat U\in\mathbb{C}^{2^m\times 2^m}$ into the 5-qubit space by padding with identity,
\begin{equation}\label{eq:padding}
\mat U_{\mathrm{pad}} \;=\; \mat U \otimes \mat I_{2^{5-m}},    
\end{equation}
which corresponds to acting trivially on the additional qubits. Notably, this simple embedding works despite the fact that the model is never trained on padded examples. We emphasize here again that prior learned approaches typically train separate models for each qubit count, while we train only a single model.

\subsection{Evaluation on Synthetic Data}

We evaluate the MDL-guided synthesis on randomly generated Clifford+$\T$ targets drawn using the rejection-sampling method described in~\cref{p:data}, following the protocol of~\cite{rietsch2024unitary} for comparability:
we sample a target $\T$-count and generate a random Clifford+$\T$ circuit whose peephole-optimized form preserves that $\T$-count. Unless stated otherwise, we sweep $\T$-counts from $0$ to $20$ and consider circuits with gate counts in the range $[3,60]$. For each $\T$-count we generate $100$ test instances and report the number of successful syntheses under a fixed compute budget.

We verify synthesis correctness using the phase-invariant average gate fidelity $F_{\mathrm{avg}}$ in~\cref{eq:AvgFidelity}. We count a run as successful if the synthesized circuit unitary $\mat U(C)$ satisfies $F_{\mathrm{avg}}(\mat U(C),\mat U^\star)\ge 0.9$. We choose $0.9$ for comparability as this is the threshold used in~\cite{rietsch2024unitary}. Unless stated otherwise, we use beam width $B{=}10$ and a budget of $8000$ trials (\cref{sec:exploration}) per instance. This results in an average runtime of $\sim$22 seconds per instance on an \textsc{NVIDIA A100-80GB} GPU.

\begin{figure*}[t]
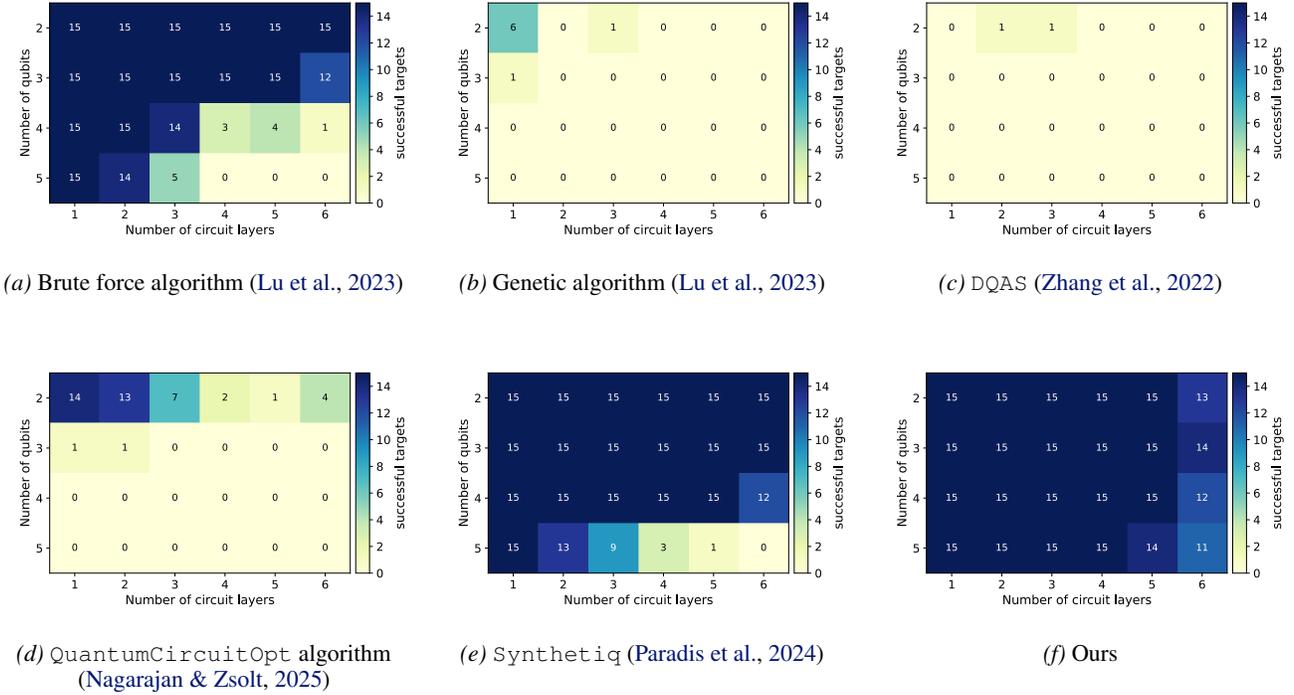

  \centering
  \begin{subfigure}[t]{0.32\textwidth}\centering
     \includesvg[width=\linewidth]{gfx/QASHeatmaps/qas_success_heatmap_search}
    \caption[Brute force algorithm]{Brute force algorithm \cite{lu2023qasbench}}
  \end{subfigure}\hfill
  \begin{subfigure}[t]{0.32\textwidth}\centering
    \includesvg[width=\linewidth]{gfx/QASHeatmaps/qas_success_heatmap_genetic}
    \caption[Genetic algorithm]{Genetic algorithm \cite{lu2023qasbench}}
  \end{subfigure}\hfill
  \begin{subfigure}[t]{0.32\textwidth}\centering
    \includesvg[width=\linewidth]{gfx/QASHeatmaps/qas_success_heatmap_dqas}
    \caption{\texttt{DQAS} \cite{zhang2022dqas}}
  \end{subfigure}\hfill
  
  \vspace{0.6em}

  \begin{subfigure}[t]{0.32\textwidth}\centering
    \includesvg[width=\linewidth]{gfx/QASHeatmaps/qas_success_heatmap_qopt}
    \caption[QuantumCircuitOpt algorithm]{\texttt{QuantumCircuitOpt} algorithm \cite{nagarajan2025provably}}
  \end{subfigure}\hfill
  \begin{subfigure}[t]{0.32\textwidth}\centering
    \includesvg[width=\linewidth]{gfx/QASHeatmaps/qas_success_heatmap_synthetiq}
    \caption{\texttt{Synthetiq} \cite{paradis2024synthetiq}}
  \end{subfigure}\hfill
  \begin{subfigure}[t]{0.32\textwidth}\centering
    \includesvg[width=\linewidth]{gfx/QASHeatmaps/qas_success_heatmap_beam_099}
    \caption{Ours}
  \end{subfigure}

  \caption[QAS-Bench heatmap results]{QAS-Bench \cite{lu2023qasbench} QC Regeneration results as heatmaps for six methods. Columns correspond to layer difficulty (1--6) and rows to qubit count (2--5). Each cell shows the number of successful syntheses out of 15 targets (5 RC-S + 10 RC-C) for that $(n,\text{layer})$ bucket. All methods are re-run on the same targets under a budget-controlled per-instance wall-clock budget (22\,s for ours, 30\,s for brute force and \texttt{Synthetiq}, 60\,s for all others); darker cells indicate higher success. Our method uses a single $n{=}5$ MDL predictor with padding for $n<5$, beam width $B{=}10$ and the success criterion $F_{\mathrm{avg}}(\mat U(C),\mat U^\star)\ge 0.99$.}
  \label{fig:heatmaps}
\end{figure*}

\cref{fig:Comparison} reports synthesis success rate for different numbers of $\T$-count for 4 and 5-qubit instances comparing our approach to RL~\cite{rietsch2024unitary} and a simulated annealing algorithm \texttt{Synthetiq}~\cite{paradis2024synthetiq}. All methods saturate on low $\T$ targets, but performance diverges sharply as $\T$-count increases. In both settings, our approach degrades substantially more slowly: while the RL baseline achieves high success rates for low-to-mid $\T$ regime, it drops off rapidly.
The main reason for this is most likely as stated in \cite{rietsch2024unitary}---the training of high $\T$ regime is underrepresented during the training process. That is because training the RL model has very high computational cost.

While \texttt{Synthetiq} fails on most higher $\T$-count instances, our method maintains a high success rate deep into the hard regime. The gap widens further on 5 qubits, indicating that our approach is substantially more scalable than \texttt{Synthetiq}.
Since our method takes $\sim$ 22 seconds, we gave \texttt{Synthetiq} a timeout of 30 seconds.

\textbf{Baseline Note.} 
The comparison between our approach and \texttt{Synthetiq} is fair, but we could not re-run the RL baselines from~\cite{rietsch2024unitary} because no code was released, so we rely on the RL success counts reported in their paper; the RL success rates in~\cref{fig:Comparison} should therefore be treated as approximations. Moreover, they did not report per-instance runtime, so we consider the comparison indicative and focus the remaining experiments on controlled ablations and compute-matched baselines that we can run directly.

\subsection{Standardized Evaluation on QAS-Bench}\label{sec:qasbench}

To move beyond our synthetic data generation and enable standardized comparison, we evaluate on the QC Regeneration (circuit resynthesis) benchmark in QAS-Bench~\cite{lu2023qasbench}. Each instance specifies a target unitary and a candidate gate set; the goal is to recover an equivalent circuit over the same gate alphabet. QAS-Bench provides two folds with closely related gate sets: RandomCircuit-Single (RC-S) uses $\{\H,\S,\T,\I\}$, while RandomCircuit-Clifford (RC-C) uses $\{\H,\S,\T,\I,\mathrm{CNOT}\}$~\cite{lu2023qasbench}. Both align with our Clifford+$\T$ action set (up to the identity), with RC-S forming a strict subset that removes entangling operations.

We evaluate our model across QAS-Bench instances with $n\in\{2,3,4,5\}$, again using the same padding scheme as before (\cref{eq:padding}). QC Regeneration is organized by qubit count and circuit layer depth from 1 to 6. For each $(n,\text{layer})$ bucket, QAS-Bench includes 5 RC-S and 10 RC-C targets (15 total)~\cite{lu2023qasbench}. Because this yields only $6\times 15=90$ targets per qubit count, we aggregate RC-S and RC-C and report success counts per $(n,\text{layer})$ bucket. This aggregation is conservative for our method: on RC-S targets we still allow $\mathrm{CNOT}$ actions, enlarging the search space rather than simplifying the task. We count a synthesis as successful under the stricter criterion: $F_{\mathrm{avg}}(\mat U(C),\mat U^\star)\ge 0.99$.

We follow QAS-Bench's heatmap visualization for direct comparability (\cref{fig:heatmaps}).
For baselines, we re-run the QAS-Bench reference implementations (bi-directional brute force search and genetic programming)~\cite{lu2023qasbench} and additionally include \texttt{Synthetiq}~\cite{paradis2024synthetiq}, a differentiable quantum architecture search (\texttt{DQAS})-style baseline that relaxes discrete gate choices into continuous weights over our gate alphabet and optimizes architecture weights by gradient descent~\cite{zhang2022dqas} and a provably optimal solver \texttt{QuantumCircuitOpt} that casts circuit synthesis as a discrete mathematical optimization problem (mixed-integer programming) and returns solutions with optimality guarantees when feasible~\cite{nagarajan2025provably}.

To keep comparisons budget-controlled to our per-instance runtime ($\approx$22\,s), we use wall-clock budgets of 30\,s per instance for bi-directional brute force and \texttt{Synthetiq} and 60\,s for genetic programming, \texttt{DQAS} and \texttt{QuantumCircuitOpt} to account for higher per-iteration overhead.

As shown in~\cref{fig:heatmaps}, classical baselines degrade quickly with problem size: bi-directional brute force and \texttt{Synthetiq} solve most small instances but fail in the harder multi-qubit settings (notably $n\in\{4,5\}$ at layers $\ge 4$), while genetic programming and our \texttt{DQAS}-style baseline solve only a small fraction in the 2-qubit regime. In contrast, our MDL-guided beam search achieves 15/15 successes in almost every $(n,\text{layer})$ bucket, despite using a single predictor trained on $n{=}5$ and applied to smaller $n$ via padding.

\begin{table*}[ht]
\caption[Comparison runtime and solution size]{Runtime and solution size (lower is better) comparison on common structured circuits. $^\dagger$\textsc{Ours} uses a fixed compute budget of 8000 trials per instance (reported time is 22\,s for that budget). For other methods, time is wall-clock to produce the reported solution. \timeout\ indicates exceeding the time limit (1\,h).
The circuits are explicitly given in \cref{sec:circuits}.}
\label{tab:structured_circits}
\centering
\small
\setlength{\tabcolsep}{4pt}
\renewcommand{\arraystretch}{1.10}

\begin{tabular}{lcccccccc
  % S[table-format=2.0] S[table-format=2.0]
  % S[table-format=3.2] S[table-format=2.0]
  % S[table-format=3.2] S[table-format=2.0]
  % S[table-format=3.2] S[table-format=2.0]
}
\toprule
& \multicolumn{2}{c}{\textsc{Ours}}
& \multicolumn{2}{c}{\textsc{Synthetiq}}
& \multicolumn{2}{c}{\textsc{Brute force}}
& \multicolumn{2}{c}{\textsc{QuantumCircuitOpt}} \\
\cmidrule(lr){2-3}\cmidrule(lr){4-5}\cmidrule(lr){6-7}\cmidrule(lr){8-9}

\textbf{Problem}
& {\textbf{budget (s)}$^\dagger$} & {\textbf{gates}}
& {\textbf{time (s)}}            & {\textbf{gates}}
& {\textbf{time (s)}}            & {\textbf{gates}}
& {\textbf{time (s)}}            & {\textbf{gates}} \\
\midrule

4-GHZ state
& 22 & {\best{4}}
& 1.10 & 21
& {\best{0.09}} & {\best{4}}
& \multicolumn{2}{c}{\timeout} \\

4-linear cluster state
& 22 & {\best{7}}
& 0.93 & 20
& 3.59 & {\best{7}}
& \multicolumn{2}{c}{\timeout} \\

4-phase gadget
& 22 & {\best{7}}
& {\best{1.22}} & 32
& 28.26 & {\best{7}}
& \multicolumn{2}{c}{\timeout} \\

\midrule

5-GHZ state
& 22 & {\best{5}}
& 3.82 & 28
& {\best{1.05}} & {\best{5}}
& \multicolumn{2}{c}{\timeout} \\

5-linear cluster state
& {\best{22}} & {\best{9}}
& 132.49 & 23
& 331.26 & {\best{9}}
& \multicolumn{2}{c}{\timeout} \\

5-phase gadget
& {\best{22}} & {\best{9}}
& 224.47 & 49
& \multicolumn{1}{c}{$>742.02$} & \multicolumn{1}{c}{\na}
& \multicolumn{2}{c}{\timeout} \\

$\left[\left[5,1,3\right]\right]$-perfect code
& {\best{22}} & {\best{14}}
& 77.23 & 37
& \multicolumn{1}{c}{$>561.81$} & \multicolumn{1}{c}{\na}
& \multicolumn{2}{c}{\timeout} \\

\bottomrule
\end{tabular}
\end{table*}

To further analyze speed and solution quality of our method, we compare on known and common 4 and 5 qubit circuits in~\cref{tab:structured_circits}. 
Our approach consistently returns the smallest circuits and where brute force is feasible, matches its optimum. \texttt{Synthetiq} produces substantially larger solutions (20-32 gates) despite running in $\approx$1\,s. Unfortunately, it cannot directly incorporate the gate minimization task into its objective function, hence it finds solutions very fast that are sub-optimal. At 5 qubits, brute force becomes prohibitively expensive, whereas ours maintains the same 22\,s budget and yields compact circuits. \texttt{Synthetiq} is both slower and less efficient on the harder tasks, again showing that its scaling is limited. \texttt{QuantumCircuitOpt} fails to solve any instance within the 1\,h limit (all \textsc{timeout}).

\section{Limitations}
\textbf{Scope and scalability.}
All exact-synthesis methods we compare against---including ours---operate on \emph{dense} \(n\)-qubit unitaries and search states represented explicitly as matrices in \(\mathbb{C}^{2^n \times 2^n}\). This choice incurs an unavoidable representation cost: storing a single state already requires \(\Theta(4^n)\) complex numbers, and any state expansion that reads/writes full matrices inherits \(\Theta(4^n)\) memory traffic and arithmetic up to constant factors. Consequently, the practical \(n\) range of \emph{any} dense-matrix exact-synthesis pipeline is limited, independent of whether the heuristic is learned or solver-based. We therefore evaluate up to \(n \le 5\), matching the largest regime where the strongest prior learned and exact baselines report results (or run reliably) under comparable compute and memory budgets.
Within this shared dense-matrix setting, our focus is not to change the exponential dependence on \(n\), but to improve efficiency \emph{at fixed \(n\)} by reducing avoidable overhead. 

\textbf{Other limitations}.
As with other learned heuristics, performance depends on the training distribution and can degrade on targets that are far outside it; beam search introduces a tunable compute--quality trade-off; and the method does not provide worst-case optimality guarantees.

\section{Conclusion}

We introduced a learning-based approach to quantum unitary synthesis that uses a lightweight model to approximate the minimum description length (MDL) of candidate Clifford+T circuits and guides synthesis via stochastic beam search. The estimator is trained with supervised learning on synthesis data and is reusable across instances, enabling zero-shot inference at test time without per-instance fine-tuning or additional environment interaction. Our results show that a single trained model provides a strong, reusable heuristic: on challenging targets with high 
$\T$-count, it achieves higher success rates than state-of-the-art RL-based approaches and outperforms classical baselines in wall-clock synthesis time, improving both solution quality and runtime.

More broadly, this work highlights the value of fast, learned heuristics for scalable unitary synthesis. Unlike hand-crafted heuristics (e.g., those derived analytically in prior work such as \cite{mosca2021polynomial}), our heuristic is discovered automatically from data and can be deployed at scale with a standard search procedure. This combination---learned scoring plus efficient search---appears to be a practical path to improving synthesis performance on complex circuits where exhaustive or heavily structured methods become costly.

%Promising directions include (i) combining multiple learned and/or analytic heuristics (e.g., boosting/mixture-of-experts) to improve robustness and coverage and (ii) injecting algebraic structure of the Clifford+T gate set into the model or search---e.g., structure-aware representations or constraints---to improve generalization and interpretability. Overall, our results suggest that learned estimators meaningfully advance unitary synthesis for high-$\T$ regimes relevant to fault-tolerant quantum computing./
\section*{Impact Statement}

This paper presents work whose goal is to advance the field of machine learning. There are many potential societal consequences of our work, none of which we feel must be specifically highlighted here.

%We demonstrate that fast heuristics are a key bottleneck for scalable quantum unitary synthesis. Our method learns an effective heuristic from data and integrates it into a synthesis pipeline that scales to larger, more challenging Clifford+T instances. Across our experimental suite, this learned heuristic consistently improves performance over existing baselines (runtime and solution quality), enabling synthesis of circuits that are difficult for prior approaches.
%Unlike hand-crafted heuristics such as those used in \cite{mosca2021polynomial}, our heuristic is learned automatically, reducing reliance on manual design while retaining compatibility with large-scale search.
%Beyond the immediate gains, our results suggest several prospective research directions: combining multiple learned heuristics (e.g., boosting or mixture-of-experts) to improve robustness across circuit families and incorporating algebraic structure of the Clifford+T group into the model to further improve generalization.

\bibliography{lit.bib}
\bibliographystyle{icml2026}

%%%%%%%%%%%%%%%%%%%%%%%%%%%%%%%%%%%%%%%%%%%%%%%%%%%%%%%%%%%%%%%%%%%%%%%%%%%%%%%
%%%%%%%%%%%%%%%%%%%%%%%%%%%%%%%%%%%%%%%%%%%%%%%%%%%%%%%%%%%%%%%%%%%%%%%%%%%%%%%
% APPENDIX
%%%%%%%%%%%%%%%%%%%%%%%%%%%%%%%%%%%%%%%%%%%%%%%%%%%%%%%%%%%%%%%%%%%%%%%%%%%%%%%
%%%%%%%%%%%%%%%%%%%%%%%%%%%%%%%%%%%%%%%%%%%%%%%%%%%%%%%%%%%%%%%%%%%%%%%%%%%%%%%
\newpage
\appendix

\appendix
\onecolumn 

\section{A Transformer-based Alternative for the MDL-Predictor}
\label{sec:transformer}
While we used a simple MLP in our final configuration, we also investigated if using a transformer-architecture \cite{vaswani2017attention} as the MDL-predictor yield any improvements. This approach involved several steps: 

\subsection{Input Tokenization and Embedding}
            The model takes a unitary matrix $U \in \mathbb{C}^{2^n \times 2^n}$ as input. Each complex number in the matrix is treated as an individual token, which is processed through a multi-step procedure.
            \begin{enumerate}
                \item \textbf{Decomposition:} The input unitary $U$ is split into its real and imaginary components, forming two real-valued matrices, $U_{re}$ and $U_{im}$.
                
                \item \textbf{Floating-Point Tokenization:} Each real number in both matrices is tokenized using a method similar to MDLformer \cite{yu2025mdlformer}. A number is decomposed into a triplet of (sign, mantissa, exponent) based on a base-10 representation. For example, the value $-54.32$ would be tokenized into the triplet $(-, 5.432, \text{E}+1)$. Each part of the triplet is then mapped to a discrete token from a shared vocabulary $\mathcal{V}$:
                \begin{itemize}
                    \item Sign: `+`, `-`
                    \item Mantissa: e.g., `N0000` to `N9999` (for 4-digit precision)
                    \item Exponent: e.g., `E-100` to `E+100`
                \end{itemize}
                This process converts each complex entry $z_{ij} = x_{ij} + i y_{ij}$ into a set of 6 discrete tokens (3 for the real part, 3 for the imaginary part).
            
                 The 6 tokens for each matrix entry are converted to indices and embedded into dense vectors of dimension $d_{input}$. These 6 vectors are then concatenated and linearly projected to the model's main dimension, $d_{model}$, creating the final embedded representation grid $E \in \mathbb{R}^{2^n \times 2^n \times d_{model}}$.
                \begin{align}
                    v_{ij} &= \text{Concat}(\text{Emb}(\text{tokens}(x_{ij})), \text{Emb}(\text{tokens}(y_{ij}))) \in \mathbb{R}^{6 \times d_{input}} \\
                    E_{ij} &= \text{Linear}(\text{Flatten}(v_{ij})) \in \mathbb{R}^{d_{model}}
                \end{align}

                \item \textbf{Alternative Tokenization in $\mathbb{Z}[\tfrac{1}{\sqrt{2}},i]$:}
                It was shown in \cite{giles2013exact} that the entries of the unitary matrices in the Clifford+T gate set can be expressed in the ring $\mathbb{Z}[\tfrac{1}{\sqrt{2}},i]$. Instead of working with floating-point approximations, we can therefore express the matrix entries as follows:
                \begin{equation}
                    z_{i,j} = \tfrac{1}{2^{k_1}}\left( a_{i,j}+b_{i,j}\cdot \sqrt{2}\right) + \tfrac{1}{2^{k_2}}\left((c_{i,j}+d_{i,j}\cdot \sqrt{2})\cdot i \right)
                \end{equation}
                where $(a_{i,j}, b_{i,j}, c_{i,j}, d_{i,j}) \in \mathbb{Z}^4$ and $(k_1, k_2) \in \mathbb{N}^2$.
                We empirically found that the majority of elements can be bounded by $-\mathcal{Z} \leq a_{i,j},b_{i,j},c_{i,j},d_{i,j} \leq \mathcal{Z}$ and $0 \leq k_1,k_2 \leq \mathcal{K}$. We therefore one-hot encode $a,b,c,d$ as vectors $v_{|a|} \in \{0,1\}^{\mathcal{Z}+1}$ and $v_{\text{sign}(a)} \in \{0,1\}$ and $k$ as vectors $v_k \in \{0,1\}^{\mathcal{K}+1}$.
                We then embed these vectors using 
                \begin{equation}
                    d_{v_{|a|}} = \phi_{v_{|a|}}(v_{|a|}) \in \mathbb{R}^{d_\text{model}}\,, d_{v_{\text{sign}(a)}} = \phi_{v_{\text{sign}(a)}}(v_{\text{sign}(a)}) \in \mathbb{R}^{d_\text{model}}\,, \textrm{and}\, d_k = \phi_k(v_k)\in \mathbb{R}^{d_\text{model}}\,.
                \end{equation} 
                Afterwards we compute
                \begin{equation}
                    v_{i,j} = \text{Linear}\left(\text{Concat}\left(d_{k_1}, d_{v_{\text{sign}(a)}}, d_{v_{|a|}}, d_{v_{\text{sign}(b)}}, d_{v_{|b|}}, d_{k_2}, d_{v_{\text{sign}(c)}}, d_{v_{|c|}}, d_{v_{\text{sign}(d)}}, d_{v_{|d|}}  \right) \right) \in \mathbb{R}^{d_\text{model}}\,.
                \end{equation}
                For the elements that fall outside of these bounds we use the learnable embedding $v_\text{NaN} \in \mathbb{R}^d_\text{model}$.
                
                \item \textbf{2D Sinusoidal Positional Encoding:} To provide the model with spatial information, we use a static (non-learnable) 2D sinusoidal positional encoding, similar to the classical transformer \cite{vaswani2017attention} $PE \in \mathbb{R}^{2^n \times 2^n \times d_{model}}$. This is added to the embedded representation.
                \begin{equation}
                    H^{(0)} = E + PE
                \end{equation}
            \end{enumerate}
            
            \subsection{Axial Attention Encoder}
            The core of the model is a stack of $L$ axial attention blocks. Each block consists of a standard Transformer Encoder Layer (which includes a multi-head self-attention mechanism followed by a position-wise feed-forward network) that is applied sequentially, first along the columns and then along the rows of the input grid. This allows the model to efficiently mix information across both spatial dimensions.
            
            A single block transforms an input $H^{(l-1)}$ to $H^{(l)}$ as follows:
            \begin{enumerate}
                \item \textbf{Column-wise Processing:} The model first applies the Transformer Encoder Layer along the columns of the input tensor. This is achieved by reshaping the tensor so that the columns become the sequence dimension, applying the encoder layer, and then reshaping the result back to its original grid structure. This produces an intermediate representation, $H_{interim}$.
                
                \item \textbf{Row-wise Processing:} The intermediate tensor $H_{interim}$ is then processed along its rows. This involves a second application of the same Transformer Encoder Layer, this time with the rows treated as the sequence dimension, to produce the final output of the block, $H^{(l)}$.
            \end{enumerate}
            This sequence of operations is repeated for $L$ blocks.
            
            \subsection{Pooling and Prediction Head}
            After the final encoder block, the resulting tensor $H^{(L)} \in \mathbb{R}^{2^n \times 2^n \times d_{model}}$ contains a rich, context-aware representation of the input unitary.
            \begin{enumerate}
                \item \textbf{Flattening:} The 2D grid of representations is flattened into a single sequence of length $N_{tokens} = (2^n)^2$.
                \begin{equation}
                    H_{flat} = \text{Flatten}(H^{(L)}) \in \mathbb{R}^{N_{tokens} \times d_{model}}
                \end{equation}
                \item \textbf{Attention Pooling:} To create a weighted summary of the token representations, we use attention pooling. A learnable query vector $w_{pool} \in \mathbb{R}^{d_{model}}$ computes attention scores.
                \begin{align}
                    \alpha_k &= \frac{\exp(H_{flat, k} \cdot w_{pool})}{\sum_{j=1}^{N_{tokens}} \exp(H_{flat, j} \cdot w_{pool})} \\
                    H_{pooled} &= \sum_{k=1}^{N_{tokens}} \alpha_k H_{flat, k} \quad \in \mathbb{R}^{d_{model}}
                \end{align}
                \item \textbf{Readout Head:} A final MLP takes the pooled vector and regresses it to a single scalar value, which is the predicted circuit gate count.
                \begin{equation}
                    \widehat{CD} = \text{MLP}_{\text{out}}(H_{pooled}) \in \mathbb{R}
                \end{equation}
            \end{enumerate}
            The entire architecture is trained end-to-end using a Mean Squared Error (MSE) loss between the predicted gate count $\widehat{CD}$ and the true optimal circuit gate count.
            \begin{equation}
                \mathcal{L} = ( \widehat{CD} - CD_{true} )^2
            \end{equation}

\subsection{Comparison of Tokenization}
We found that the more powerful transformer architecture does not yield to noticeably better results and therefore does not justify the extra computational cost during inference and training (see \cref{tab:model_comparison}).

\begin{table}[t]
\centering
\caption{Comparison of three models on the validation set. Lower is better for Loss and MAE; higher is better for $R^2$.}
\label{tab:model_comparison}
\begin{tabular}{lccc}
\toprule
Model & Val.\ loss $\downarrow$ & MAE $\downarrow$ & $R^2$ $\uparrow$ \\
\midrule
Float-tokenizer & 241.8 & 12.1 & \best{0.6} \\
Alternative-tokenizer & 251.0 & 12.3 & \best{0.6} \\
MLP & \best{67.1} & \best{7.7} & 0.55 \\
\bottomrule
\end{tabular}
\end{table}

\newpage

\section{Training and Hyperparameter Details}

\paragraph{Baseline MLP (MDL predictor).}
\begin{itemize}
  \item \textbf{Data generation.} We sample random Clifford+$T$ circuits using rejection sampling over $T$-counts, compute the corresponding unitaries, apply a greedy peephole optimizer and use the \emph{optimized circuit gate count} as the regression target. For longer circuits, we add intermediate training examples by truncating the circuit after half and three-quarters of the $T$ gates (curriculum-style augmentation).
  \item \textbf{Architecture.} We train a 5-qubit MLP with hidden layer widths $\{1024,512,128\}$ and dropout $0.0$. 
  \item \textbf{Optimization.} We use AdamW ($\beta_1=0.9$, $\beta_2=0.999$, $\epsilon=10^{-8}$) with mean-squared error (MSE) loss, weight decay $0.0$, and gradient clipping at $1.0$.
  \item \textbf{Training schedule.} The per-step batch size is 256 with 5-step gradient accumulation (effective batch size 1280). We train for 200 epochs with 2000 steps per epoch. We use a 2000-step warmup followed by cosine learning-rate decay with $T_{\max}=20000$ and $\eta_{\min}=10^{-6}$; the peak learning rate is $5\times 10^{-4}$.
  \item \textbf{Data ranges and throughput.} Training covers gate counts in $[1,60]$ and $T$-counts in $[0,20]$. We use a replay buffer of size 6000 and a validation set of size 4096, with 29 dataloader workers and 200 validation steps per epoch. End-to-end training takes $\approx$6 hours on 30 CPU cores with a 4\,GB GPU.
\end{itemize}

\paragraph{CD-Former (transformer model).}
\begin{itemize}
  \item \textbf{Inputs and target.} Unitaries are tokenized using the float tokenizer (mantissa precision 4; exponent range $[-10,10]$) or the alternative tokenizer (see \cref{sec:transformer}). The supervision target is the \emph{remaining} Clifford+$T$ gate count.
  \item \textbf{Architecture.} We use a 5-qubit Transformer encoder with model width 96, 4 attention heads, 8 encoder layers, feedforward dimension 384, dropout 0.1, and embedding dimension 96.
  \item \textbf{Optimization.} We use AdamW ($\beta_1=0.9$, $\beta_2=0.999$, $\epsilon=10^{-8}$) with MSE loss, weight decay 0.01 and gradient clipping at $5.0$.
  \item \textbf{Training schedule.} The per-step batch size is 240 with 5-step gradient accumulation (effective batch size 1200). We train for 100 epochs with 10k steps per epoch. We use a 50k-step warmup followed by cosine decay with $T_{\max}=10^{6}$ and $\eta_{\min}=10^{-5}$; the peak learning rate is $2\times 10^{-4}$.
  \item \textbf{Data ranges and throughput.} Training covers gate counts in $[1,60]$ and $T$-counts in $[0,20]$. We use a replay buffer of size 4000 and a validation set of size 4096, with 29 dataloader workers and 200 validation steps per epoch.
\end{itemize}

\section{Structured circuits}
\label{sec:circuits}

\begin{figure}[ht]
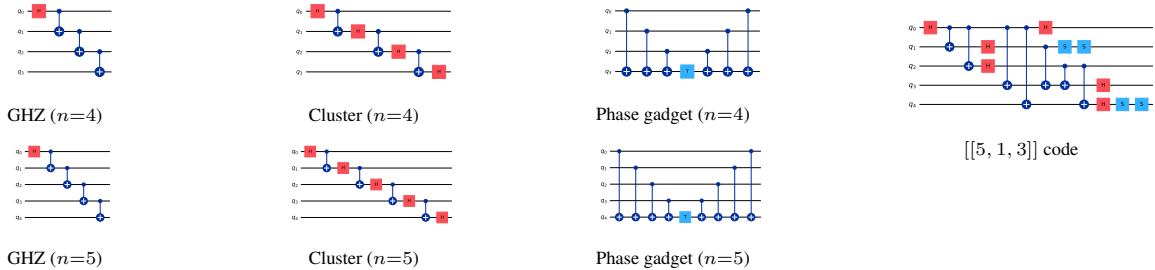

\centering

% Left: 3x2 grid of circuits
\begin{minipage}[t]{0.72\textwidth}
\centering
\setlength{\tabcolsep}{4pt}   % horizontal padding between columns
\renewcommand{\arraystretch}{1.0}

\begin{tabular}{ccc}
  \begin{minipage}[c]{0.31\linewidth}\centering
    \includesvg[width=\linewidth]{gfx/structured_circuits/ghz_4_circuit}\\[-0.3em]
    {\scriptsize GHZ ($n{=}4$)}
  \end{minipage}
&
  \begin{minipage}[c]{0.31\linewidth}\centering
    \includesvg[width=\linewidth]{gfx/structured_circuits/cluster_4_circuit}\\[-0.3em]
    {\scriptsize Cluster ($n{=}4$)}
  \end{minipage}
&
  \begin{minipage}[c]{0.31\linewidth}\centering
    \includesvg[width=\linewidth]{gfx/structured_circuits/phase_gadget_4_circuit}\\[-0.3em]
    {\scriptsize Phase gadget ($n{=}4$)}
  \end{minipage}
\\[2mm]
  \begin{minipage}[c]{0.31\linewidth}\centering
    \includesvg[width=\linewidth]{gfx/structured_circuits/ghz_5_circuit}\\[-0.3em]
    {\scriptsize GHZ ($n{=}5$)}
  \end{minipage}
&
  \begin{minipage}[c]{0.31\linewidth}\centering
    \includesvg[width=\linewidth]{gfx/structured_circuits/cluster_5_circuit}\\[-0.3em]
    {\scriptsize Cluster ($n{=}5$)}
  \end{minipage}
&
  \begin{minipage}[c]{0.31\linewidth}\centering
    \includesvg[width=\linewidth]{gfx/structured_circuits/phase_gadget_5_circuit}\\[-0.3em]
    {\scriptsize Phase gadget ($n{=}5$)}
  \end{minipage}
\end{tabular}
\end{minipage}
\hfill
% Right: last circuit alone
\begin{minipage}[t]{0.26\textwidth}
\centering
\includesvg[width=\linewidth]{gfx/structured_circuits/perfect_code_circuit}\\[-0.3em]
{\scriptsize $[[5,1,3]]$ code}
\end{minipage}

\caption{Canonical structured circuits used in our evaluation: GHZ states, cluster states, phase-gadget constructions and the $[[5,1,3]]$ (perfect) quantum error-correcting code.}
\label{fig:structured-circuits}
\end{figure}

This appendix collects the structured circuit families used throughout the experiments: GHZ states~\cite{Greenberger1989ghz}, cluster states~\cite{Raussendorf2001onewayqc}, phase gadgets as formalized in the ZX-calculus~\cite{vandewetering2020zxcalculus}, and the $[[5,1,3]]$ perfect code~\cite{Laflamme1996perfectcode}. These circuits (shown in~\cref{fig:structured-circuits}) serve as standardized targets with known structure, enabling controlled comparisons across qubit counts and circuit motifs.

\section{Effect of exploration}
\label{sec:exploration}
A trial budget of 4,000 is the largest setting that fits comfortably on an 80 GB GPU using a $n=5$ qubit model; 8,000 trials is exactly a $2\times$ increase. Accordingly, we implement an 8,000-trial evaluation by running two consecutive 4,000-trial calls, for a total runtime of $\sim$22\,s. With two GPUs, these two 4,000-trial calls can be run in parallel, reducing the runtime to $\sim$11\,s.

\begin{figure}[ht]
	\centering
	\includesvg[width=\linewidth,height=0.25\textheight,keepaspectratio]{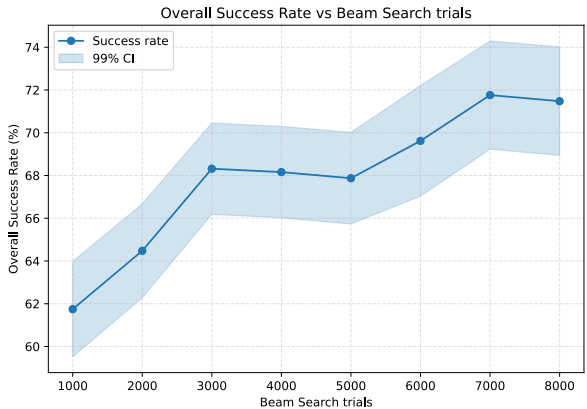}
    \caption{Overall success rate as a function of the beam-search trial budget.}
	\label{fig:trials-vs-success}
\end{figure}

For this isolated exploration ablation we additionally train an $n=4$ predictor to remove padding confounds; all main results use the single $n=5$ predictor. To assess the effect of exploration, we vary the number of trials per circuit on the $n=4$ qubit setting using an MLP trained on $n=4$ qubits (i.e., no padding). Here 8000 trials can be run on a single GPU because the $n=4$ model is smaller than the $n=5$ model. For each time step $t\in\{0,\dots,20\}$ (same protocol as in~\cref{sec:experiments}), we measure the fraction of solved circuits and compare performance across trial budgets. Results are shown in \cref{fig:trials-vs-success}.

\cref{fig:trials-vs-success} plots the mean overall success rate (solid line) as a function of the beam-search trial budget, aggregated over the evaluation protocol described above. Performance increases monotonically with additional trials, with the largest gains occurring when moving from low budgets (roughly 1{,}000--3{,}000 trials) and diminishing returns thereafter; beyond $\sim$5{,}000 trials, improvements are smaller and begin to plateau. The light-blue shaded region denotes a $99\%$ confidence interval around the mean success rate at each budget: narrower bands indicate more precise estimates, while wider bands indicate higher uncertainty.

This saturation is expected because the number of distinct symmetry transforms is bounded by $n!$ (and $\times 2$ if including inverses), so additional trials increasingly revisit previously explored transforms. Residual gains at higher budgets come primarily from stochasticity (Gumbel noise), which can introduce additional diversity even when sampling under the same symmetry.

\section*{Disclosure: Use of Generative AI Tools}
We used GPT-5.2 (OpenAI) to assist with drafting and editing the manuscript text (wording, clarity and grammar). The authors produced and verified all technical content, claims, experiments and conclusions and take full responsibility for the final manuscript.
%%%%%%%%%%%%%%%%%%%%%%%%%%%%%%%%%%%%%%%%%%%%%%%%%%%%%%%%%%%%%%%%%%%%%%%%%%%%%%%
%%%%%%%%%%%%%%%%%%%%%%%%%%%%%%%%%%%%%%%%%%%%%%%%%%%%%%%%%%%%%%%%%%%%%%%%%%%%%%%

\end{document}